\documentclass[a4paper,10pt]{article}
 \usepackage[english]{babel}
  \usepackage{amsmath,amsfonts,amssymb,amsthm}
   \usepackage{graphicx,xcolor}
    \usepackage{subfigure}
\newcommand{\sizedef}{
        \headheight=0pt                               
 	  \topmargin=-1.5cm \headsep=1.5cm              
        \textheight=21truecm \textwidth=13truecm    
 	  \setlength{\columnsep}{20pt}                  
  }
\sizedef

\def\Z{\mathbb{Z}}

\def\E{\mathcal E}

\let\e=\epsilon
\let\d=\delta

\newcommand{\be}{\begin{equation}}
\newcommand{\ee}{\end{equation}}
\newcommand{\ba}{\begin{eqnarray}}
\newcommand{\ea}{\end{eqnarray}}
\newcommand{\nn}{\nonumber \\}
\newcommand{\finedim} {\begin{flushright}
                         $\Box$
                        \end{flushright}}

\def\bbm[#1]{\mbox{\boldmath $#1$}}

\newtheorem{prop}{Proposition}

\newtheorem{oss}{Remark}
\newtheorem{Def}{Definition}

\begin{document}

\title{Equivariant singularity analysis of the 2:2 resonance}

\author{Antonella Marchesiello \\
Dipartimento di Scienze di Base e Applicate per l'Ingegneria\\
Universit\`a di Roma ``la Sapienza"\\
Via Antonio Scarpa, 16 - 00161 Roma\\
\\
and\\
\\
Giuseppe Pucacco\thanks{pucacco@roma2.infn.it.}\\
Dipartimento di Fisica and INFN -- Sezione di Roma II\\
Universit\`a di Roma ``Tor Vergata"\\
Via della Ricerca Scientifica, 1 - 00133 Roma}

\maketitle

\begin{abstract}
We present a general analysis of the bifurcation sequences of 2:2 resonant reversible Hamiltonian systems invariant under spatial $\Z_2\times\Z_2$ symmetry. The rich structure of these systems is investigated by a singularity theory approach based on the construction of a universal deformation of the detuned Birkhoff normal form. The thresholds for the bifurcations are computed as asymptotic series also in terms of physical quantities for the original system.
\end{abstract}

\section{Introduction}\label{sec:1}
We consider the problem of determining the phase-space structure of a Hamiltonian  describing a ``2:2 resonance''. With this we mean a Hamiltonian dynamical system close to an equilibrium with almost equal unperturbed positive frequencies and which is invariant with respect to reflection symmetries in both symplectic variables in addition to the time reversion symmetry. We aim at a general understanding of the bifurcation sequences of periodic orbits in general position from/to normal modes, parametrized by an internal parameter (the ``energy'') and by the physical parameters: the independent coefficients characterising the non-linear perturbation and a ``detuning'' parameter associated to the quadratic unperturbed Hamiltonian.

Among low-order resonances (see e.g. \cite{SV}) the symmetric 1:1 resonance plays a prominent role. The general treatment is attributed to Cotter \cite{Cotter} in his PhD thesis, but several other works explored its generic features \cite{Br1:1,MRS1,MRS2,vdM}. Particular emphasis has been given to the symmetric subclass which is the subject of the present paper. In particular, we recall the works of Kummer \cite{Ku}, Deprit and coworkers \cite{D1,DE,M1} and Cushman \& Rod \cite{CR}. The connection of equivariant singularity theory and bifurcation of periodic orbits was made for the first time in \cite{GB1} for $\Z_2$-equivariance and in \cite{vdM85} for $\mathbb{S}^1$-equivariance. Broer and coworkers \cite{Br1:1} exploit equivariant singularity theory with distinguished parameters to study resonant Hamiltonian systems. We proceed on the same ground to detail the application of an equivariant singularity analysis to the generic unfolding of a detuned 1:1 resonance invariant under $\Z_2 \times \Z_2$ mirror symmetries in space and reversion symmetry in time.

Among several areas of application in physics, chemistry and engineering, great relevance plays the application of resonance crossing to galactic dynamics \cite{zm,noiApJ}; recent treatments have been given in \cite{MP11,MP13}. We consider systems in two degrees of freedom,
therefore, in order to classify the dynamics with singularity theory,
we need to perform a preliminary transformation by constructing a normal form of the physical source problem \cite{Cic, gior}.

After the normalization procedure, the system acquires an additional (formal) $\mathbb{S}^1$ symmetry.
Using regular reduction \cite{CB}, we divide out the $\mathbb{S}^1$ symmetry of the normal form obtaining a planar system: this allows us to apply singularity theory to get a universal unfolding \cite{Br1:1,Br1:2}.

Actually we have to respect the symmetries and reversibility of the original system, implying the invariance of the
planar system with respect to the $\Z_2 \times \Z_2$ action on $\mathbb R^2$ and
 thus we are lead in the framework of $\Z_2 \times \Z_2$-{\it equivariant} singularity theory.
The momentum corresponding  to the $\mathbb{S}^1$ symmetry serves as the internal   ``distinguished'' parameter \cite{BrL,Brc}. The planar system
can be further simplified into a versal deformation of the germ of the singularity \cite{dui}.
The basic classification proceeds by examining the inequivalent cases corresponding to the two sign combinations of the quartic terms in the germ \cite{HDS,Hlibro}. In \cite{GMSD,vdM96} it is shown that after $\mathbb{S}^1$-reduction one actually obtains $\Z_2$-invariant bifurcation equations. In the discussion in section \ref{sec:bifu} below we see the form taken by these equations in the present context.

The simplifying transformations inducing the planar system from its universal deformation are explicitly computed,
so that we are able to obtain the bifurcation sequences of the detuned 2:2 normal form. Deformation parameters are determined by the coefficients of the quartic terms: they fix the \emph{qualitative} picture whereas the inclusion of higher-order terms gives only small \emph{quantitative} effects which do not change the qualitative overall results. This allows us to pull back the bifurcation curves
to the original parameter-energy space \cite{Ku, DE, M1}.
In particular, we find out the \emph{physical} energy threshold values
(depending on the coefficients of the original system and on the detuning parameter)
which determine the pitchfork bifurcation of periodic orbits in general position (namely \emph{loop} and
\emph{inclined} orbits in the present case) from/to the normal modes of the original system.

The plan of the paper is the following: in section \ref{sec:2} we introduce the model problems and their Birkhoff normal forms; in section \ref{sec:centr_reduction} we perform the reduction to the planar system and derive its central singularity; in section \ref{sec:sing_tr} we introduce the versal deformation and describe the algorithm to induce the models from it; in section \ref{sec:bifu} we classify possible dynamics by identifying the bifurcation sequences; in section \ref{sec:original11} we discuss the implications of these results for the original physical models; in the Appendix we provide the normal forms and list the explicit values of coefficients appearing in the transformation series.

\section{The model and its normal form}\label{sec:2}

Let us consider a two-degrees of freedom system whose Hamiltonian is an analytic function in a neighborhood of an elliptic equilibrium and symmetric under reflection with respect to both symplectic variables. Its series expansion about the equilibrium point can be written as

\begin{equation} \label{Hamiltonian}
\mathcal H(\mathbf{p},\mathbf{x})=\sum_{j=0}^{\infty} \mathcal H_{2j}(\mathbf{p},\mathbf{x})
\end{equation}
where each term is a homogeneous polynomial of degree $2(j+1)$ exhibiting two $\Z_2$ symmetries,
denoted $S_1$ and $S_2$:

\begin{eqnarray}
S_1&:&(x_1,x_2,p_1,p_2)\rightarrow(-x_1,x_2,-p_1,p_2)\label{spatial_symmetry1}\\
S_2&:&(x_1,x_2,p_1,p_2)\rightarrow(x_1,-x_2,p_1,-p_2)\label{spatial_symmetry2}
\end{eqnarray}
and the time reversion symmetry
\be \label{time_symmetry}
T: (x_1,x_2,p_1,p_2)\rightarrow(x_1,x_2,-p_1,-p_2)
\ee
To take into account the presence of reflection symmetries,
we will speak of a 2:2 resonance. We remark that the Hamiltonian function \eqref{Hamiltonian} could also be invariant under  other
transformations, such as reflections acting on the $\mathbf{x}$ and not on the $\mathbf{p}$ and viceversa.
Our choice to consider  reflection symmetries \eqref{spatial_symmetry1} and \eqref{spatial_symmetry2}
lies in the Lagrangian description of a reversible system, giving up using all possibilities of the Hamiltonian description \cite{BH}.

We assume the zero-order term ${\mathcal H}_0$ to be in the positive definite form

\be
{\mathcal H}_0(\mathbf{p},\mathbf{x})=\frac12\omega_1(p_1^2+x_1^2)+\frac12\omega_2(p_2^2+x_2^2).
\ee
where the two harmonic frequencies
$\omega_1$ and $\omega_2$ are generically not commensurable. In unperturbed harmonic oscillators frequency ratios are fixed. However, the non-linear coupling between the degrees of freedom induced by the perturbation causes the frequency ratio to change. Therefore, even if the unperturbed system is non-resonant,  the system passes through resonances of order given by the integer ratios closest to the ratio of the unperturbed frequencies.
This phenomenon is responsible for the birth of new orbit families bifurcating from the normal modes or from lower-order resonances \cite{Bi,conto1,CM,Vf}. Therefore, to catch the main features of the orbital structure, it is convenient to
 assume  the frequency ratio  not far from $1$ and then approximate it by introducing a small ``detuning'' $\delta$ \cite{Vf}, so that
\be\label{det11}
\omega_1=(1+\d)\omega_2.
\ee
Hence, after a scaling of time

\be
t \longrightarrow \omega_2 t,\ee
so that
\be\label{scaling_t}
\frac1{\omega_2} {\mathcal H} \doteq H = \sum_{j=0}^{\infty} H_{2j},
\ee
the unperturbed term turns into

\be
H_0(\mathbf{p},\mathbf{x})=\frac12(p_1^2+p_2^2+x_1^2+x_2^2)
\ee
and we can construct a $2:2$ detuned normal form by proceeding as if the unperturbed harmonic part would be in exact $1$:$1$ resonance and including the remaining part, which we assume of second order, inside the perturbation.

In practice, working with formal power series the expansions are truncated at some $j_{\rm max} \doteq N$.
If we truncate the normalization procedure to the minimal  order required, i.e. $N=1$ \cite{conto},
 the system turns out to be already reduced to the universal unfolding. At order $N>1$ this is not true anymore and we need the algorithms described in sections \ref{sec:centr_reduction}--\ref{sec:sing_tr}. In the following we truncate at order $N=2$ (i.e. including terms up to the sixth degree), but the procedure can be iterated to arbitrary higher orders.
 
 For sake of clarity we consider the natural case, so that
the higher-order terms in the Hamiltonian $H$ read

\be\label{perturbation}
H_{2j} = \sum_{k=0}^{j+1}  v_{2k,2(j+1-k)} \, x_1^{2k} x_2^{2(j+1-k)},
\ee
where $v_{2k,2(j+1-k)}$, with $j \in \mathbb{N}, 0 \le k \le j+1$, are physical coefficients which may depend on the detuning \cite{CDHS} and in this respect we keep terms of this kind resulting from rescaling. The general (not natural) case can be treated in an analogous way.

Proceeding with a Birkhoff  normalization procedure \cite{Bir,Cic,gior} up to order $N$, we obtain the ``normal form''

\be
K(J_1,J_2,2\phi_1-2\phi_2) = \sum_{j=0}^{N}  K_{2j},
\label{aav11}
\ee
where we  have introduced the action-angle(--like) variables with the transformation:

\be\label{Vaa}
x_{\ell}=\sqrt{2J_{\ell}}\cos\phi_{\ell},\;\;\;\;\;p_{\ell}=\sqrt{2J_{\ell}}\sin\phi_{\ell}, \quad \ell=1,2.
\ee
This Hamiltonian is in normal form with respect to the quadratic unperturbed part $H_0$ that in these coordinates reads

\be H_0^{AA} = K_0 = J_1 + J_2 \doteq \E. \ee
We remark that for the computation of \eqref{aav11} and results thereof, the use of algebraic manipulators like  {\sc mathematica}\textregistered \ is practically indispensable: in Appendix A we report the terms up to second order of this and the transformed normalized functions. After the normalization, the system has acquired an additional $\mathbb{S}^1$ symmetry. The corresponding conserved quantity is given by $H_0 = \E$. This enables us to formally reduce \eqref{aav11}
to a planar system. It is well known \cite{Vf,zm,MP11} that, in addition to the normal modes, periodic orbits ``in general position'' may appear. They exist only above a given threshold when the normal modes suffer stability/instability changes. This phenomenon can be seen as a {\it bifurcation} of the new family from the normal mode when it enters in 1:1 resonance with a normal perturbation (or as a disappearance of the family in the normal mode). The phase between the two oscillations also plays a role. These additional periodic orbits are respectively given by the conditions $\phi_1-\phi_2=0,\pi$ ({\it inclined} orbits) and $\phi_1-\phi_2=\pm\pi/2$ ({\it loop} orbits): therefore both families give two orbits. We are going to investigate the general occurrence of these bifurcations as they are determined by the internal and external parameters. In practice we analyze the nature of critical points of an integrable approximation of the iso-energetic Poincar\'e map provided by the phase-flow of a planar system obtained through reduction and further simplification of the normal form.

\section{Reduction to the central singularity of the planar system}\label{sec:centr_reduction}

We perform the following canonical transformation \cite{Br1:1}

\be \label{suber}
\left\{
  \begin{array}{l}
    J_1=J \\
    J_2=\E-J \\
     \psi=\phi_1-\phi_2\\
    \chi=\phi_2
  \end{array}
\right.
\ee
and, since $\chi$ is cyclic and its conjugate action $\E$ is the additional integral of motion, we may introduce the effective
Hamiltonian

\be\label{KER11gen}
\widetilde {\mathcal K} (J,\psi;\E,\d)=\E+\sum_{j=1}^N \left(\mathcal A_j(J;\E,\d)+\mathcal B_j(J;\E,\d)\cos2\psi\right)
\ee
where $\mathcal{A}_j,\mathcal{B}_j$ are homogeneous polynomials listed in Appendix \ref{sec:bb} for $N=2$. We get a one degree of freedom system: in the following we refer to it as {\it the} $1$DOF system.

We now perform a further reduction into a planar system, viewing $\E$
as a \emph{distinguished} parameter \cite{Br1:2}.

\begin{oss} The adjective distinguished
refers to the fact that $\E$ stems from the phase space of $\widetilde{\mathcal K}$ and is a parameter only for the $1${\rm DOF} system, not for the original one.
\end{oss}
\noindent The planar reduction is obtained via the canonical coordinate transformation \cite{Ku}

\be\label{ccoord}
\left\{
  \begin{array}{l}
    x=\sqrt{2J}\cos\psi, \\
    y=\sqrt{2J}\sin\psi,
  \end{array}
\right.
\ee
so that the Hamiltonian function $\widetilde{\mathcal K}$
is converted into the planar Hamiltonian

\be\label{Kxy11}
\mathcal K(x,y;\E,\d) = \sum_{i=0}^3\sum_{j=0}^{3-i}c_{2i,2j}x^{2i}y^{2j}+ h.o.t.
\ee
where $c_{2i,2j}=c_{2i,2j}(\E,\d)$. The actions \eqref{spatial_symmetry1} and \eqref{spatial_symmetry2}
reduce to $(x,y)\rightarrow$ $(-x,-y)$ and the time reversion symmetry reduces to $(x,y)\rightarrow$ $(x,-y)$.
Therefore, the planar Hamiltonian turns out to be invariant under a $\mathbb{Z}_2\times \mathbb{Z}_2$ action on $\mathbb R^2$.

\begin{oss}[Singular circle]\label{oss:singular_circle}
The coordinate transformation \eqref{Vaa} is singular at the coordinate axes $J_1=0$ and
$J_2=0$. After the transformation \eqref{suber}, these axes respectively become $J=0$ and $J=\E$. The first
singularity is removed by introducing the Cartesian coordinates \eqref{ccoord} in the plane.
The second singularity is called ``singular circle'' and is
given by

\be\label{sing_circ1}
x^2+y^2=2\E.
\ee
On this circle $J_2=0$ so that the coordinate $\phi_2$ is ill defined and therefore $\psi$ is. In particular, this implies that $\widetilde{\mathcal K}$ is constant on this circle.
\end{oss}

Since the system is planar now, we may use general ($\mathbb{Z}_2\times \mathbb{Z}_2$-equivariant)
planar transformations for further reductions,  as opposed to just the canonical ones  \cite{Br1:1}.
The resulting system is not conjugate but \emph{equivalent} to the original one. At this point the system  depends on a
distinguished parameter $\E$, a detuning parameter $\d$ and several ordinary coefficients. The parameters
are supposed to be \emph{small}. We look at the degenerate Hamiltonian that results
when $\d=0$ (resonance) and $\E=0$ (the diameter of the singular circle vanishes). This is called the \emph{central singularity}, also
known as the \emph{organizing center}. 

At the singular values of the parameters we have
that $\mathcal K$ reduces to

\be
{\mathcal K}_s(x,y)\doteq\mathcal K|_{\d=0,\E=0}(x,y)=s_{4,0} x^4 +s_{2,2} x^2 y^2+s_{0,4} y^4 + h.o.t.,  \label{sgerm}
\ee
where $s_{i,j}=c_{i,j}(0,0)$. In particular

\be
s_{4,0}=A-3C,\;\;\;
s_{2,2}=2(A-2C),\;\;\;
s_{0,4}=A-C
\ee
where

\be
A=\frac38 (v_{4,0}+v_{0,4}),\;\;\;
C=\frac18 v_{2,2}.
\ee
The constant term $c_{0,0}$ can be neglected and,  by a simple scaling transformation,
${\mathcal K}_s$ can be turned into

\be\label{ssgerm}
 {\mathcal K}'_s(x,y)=\e_1 x^4+\mu x^2 y^2+\e_2 y^4 + h.o.t., \;\;\;\; \e_1,\e_2 \in \{-1,1\}
\ee
where

\be
\mu=\frac{2(A-2C)}{\sqrt{|(A-3C)(A-C)|}}, \;\;
\e_1 = \frac{A-3C}{|A-3C|}, \;\; \e_2 = \frac{A-C}{|A-C|}.
\ee

\begin{oss} [Non-degeneracy conditions]\label{oss:degeneracy}
This is possible provided that the coefficients of $x^4$ and $y^4$ in ${\mathcal K}'_s$ are not zero.
This translates into the non-degeneracy conditions

\be\label{condition:nondegeneracy}
A-3C\neq0\;\;\;\; \hbox{ and }\;\;\;\; A-C\neq0.
\ee
The sign
of $\e_1$ and $\e_2$ is determined by the sign of $A-3C$ and $A-C$ respectively. In the following, we will look for a transformation
which brings the system at the central singularity into the standard form
\eqref{11central_singularity}. If conditions \eqref{condition:nondegeneracy} are not satisfied this is not possible: a reduction
of \eqref{sgerm} may still be possible, however we may have to retain sixth degree (or even higher) order terms in the central singularity.
\end{oss}

We now start the procedure of simplifying the reduced normal form by following the so called {\it BCKV-restricted reparametrization} method \cite{Brc}. As first step in the process of obtaining the versal deformation,  we look for a near identity planar morphism $\Phi(x,y)$  which brings the system at the central singularity into the polynomial form
\be\label{11central_singularity}
f(x,y)\doteq\e_1 x^4 +\mu x^2 y^2+\e_2y^4.
\ee
This morphism has to respect the $\mathbb{Z}_2\times\mathbb{Z}_2$ symmetry $(x,y)\rightarrow(\pm x,\pm y)$.
The following proposition \cite{Br1:1} assures the existence of the transformation we are looking for:

\begin{prop} \label{prop:determinacy}
The germ $g(x,y)=\e_1 x^4+\mu x^2y^2+\e_2 y^4+$ h.o.t., with $\e_i=\pm1$, is $\mathbb{Z}_2\times\mathbb{Z}_2$ isomorphic to $f(x,y)=\e_1 x^4+\mu x^2y^2+\e_2 y^4$, provided that $\mu^2\neq4\e_1\e_2$.
\end{prop}
For our system the condition

\be\label{condition:determinacy}
\mu^2\neq4\e_1\e_2
\ee
is equivalent to require that $C\neq0$. Making this assumption, we are able
to compute $\Phi$ using the  iterative procedure described in \cite{Br1:2} here adapted to our symmetric context. We set $\Phi^{(1)}_1(x,y)=x$, $\Phi^{(2)}_1(x,y)=y$ and assume that for some $k$

 $${\mathcal K}'_s\circ\Phi_k=\e_1 x^4+\mu x^2y^2+\e_2y^4+ O(|x,y|^{2(k+2)}).$$
 Then we set

 \begin{eqnarray}
 \Phi^{(1)}_{k+1}&=&\Phi^{(1)}_k+\sum_i\alpha^{(1)}_iP^{(k+1)}_i \\
 \Phi^{(2)}_{k+1}&=&\Phi^{(2)}_k+\sum_i\alpha^{(2)}_iQ^{(k+1)}_i
 \end{eqnarray}
 where $\{ P^{(k+1)}_i\}$, $\{ Q^{(k+1)}_i\}$ respectively span the space of  two variables monomials  of degree $k+1$, invariant under the $\Z_2$ actions $(x,y)\rightarrow(x,\pm y)$ and $(x,y)\rightarrow(\pm x,y)$. The coefficients $\alpha_i^{(j)}$ are to be found in order to cancel the terms of order $O(|x,y|^{2(k+2)})$ in ${\mathcal K}'_s$. This translates
into a set of linear equations for the real numbers $\alpha_i^{(j)}$. By the existence of the reducing transformation, this set of equations is never over determined and can always be solved if \eqref{condition:determinacy} is satisfied.

If we compute $\Phi$ up to order $2$ in $k$ we get the following proposition

\begin{prop}\label{prop:Kb}
Let us consider the planar Hamiltonian ${\mathcal K}$. Except for the exceptional values $C=A$, $C=A/3$ and
$C=0$, there exists a coordinate transformation $\Phi:\mathbb{R}^2\rightarrow\mathbb{R}^2$ such that
${\mathcal K}_b\doteq{\mathcal K}\circ\Phi$ is of the form

\ba
{\mathcal K}_b(x,y; \E)&=&(a_1 \E + a_2 \E^2 + b_1 \E + b_2 ) x^2 + (a_3\E + a_4 \E^2 + b_3\E +
    b_4) y^2 \nn
&+& (\e_1 + a_5 \E + b_5)x^4 + (\mu + a_6 \E +
    b_6) x^2 y^2  \nn
&+& (\e_2 + a_7\E + b_7)y^4+ h.o.t.,\label{Kb}
\ea
where the $a_i$ are coefficients and the $b_i$ are parameters linearly depending on $\d$ and vanishing at $\d=0$.
They are listed in appendix \ref{app:11cat}.
Neglecting terms
of $O(|x,y|^5)$ the following is a suitable transformation $\Phi$:

\be
\Phi:\left\{
       \begin{array}{ll}
         x\rightarrow &x+ \frac{\e_1( 2\e_2s_{2,4}-\mu s_{0,6})\mu  +2 \mu s_{6,0}-4\e_1 s_{4,2} }{4 \e_1\e_2 \left(4\e_1\e_2-\mu^2\right)}x y^2 -\frac{s_{6,0}\e_1}{4}x^3 \\
         y\rightarrow & y+\frac{\e_2(+2\e_1s_{4,2} -\mu s_{6,0})\mu +2\mu s_{0,6}   -4 \e_2s_{2,4}}{4 \e_1\e_2\left(4\e_1\e_2-\mu^2\right)}x^2y -\frac{s_{0,6}\e_2}{4 }y^3
       \end{array}
     \right.
\ee
\end{prop}
\noindent \emph{Proof}. The existence of $\Phi$ is a consequence of proposition \ref{prop:determinacy}.
The conditions on $C$ are consequence of the non-degeneracy conditions (cfr. remark \ref{oss:degeneracy} and of condition \eqref{condition:determinacy}).
The explicit expression of the transformation up to and including terms of $O(|x,y|^3)$ has been obtained by exploiting the algorithm described above up to $k=2$.
\finedim

\section{Inducing the system from a universal deformation} \label{sec:sing_tr}

The theory assures that there exists a $\Z_2\times\Z_2$-equivariant morphism  $\phi$ which induces the reduced normal form ${\mathcal K}_b$ from a universal deformation.
In \cite{Br1:2} an algorithm is discussed in order to compute $\phi$ in presence of a $\mathbb{Z}_2$ symmetry, $(x,y)\rightarrow(x, \pm y)$. In the following, we adapt  the algorithm  to our symmetric context.

\subsection{The universal deformation}

Let us denote with $\mathcal U$ the space of all differentiable germs of two variables invariant under the action of the group

$$\Gamma=\{\rm{Id},S_1,S_2,S_1 \circ S_2\}$$
and vanishing at the origin. Moreover, let us consider the group $G$ of origin preserving $\Gamma$-equivariant $C^{\infty}$ maps on $\mathcal U$
with action on $\mathcal U$ by composition to the right. We denote by $\xi: G\times \mathcal U\rightarrow \mathcal U$ a smooth action of $G$ on $\mathcal U$.
 For a given point $f\in \mathcal U$, the action $\xi$ gives rise to an orbit,
in this notation given by $Gf$ and let $T_f(Gf)$ be the tangent space to
this orbit at the point $f$. The codimension of $T_f(Gf)$ in $T_f (U)$ is also
called the codimension of $f$. In case $f$ is given by \eqref{11central_singularity}, the codimension of $T_f(Gf)$ is finite and

 \be
 F(x,y)=\e_1x^4+(\mu+u_3)x^2y^2+\e_2y^4+u_1x^2+u_2y^2.\label{F_uni}
 \ee
is a \emph{universal deformation} of $f$ \cite{Br1:1}. This implies  that there exists a $\mathbb{Z}_2\times\mathbb{Z}_2$-equivariant
 morphism $\phi$  which induces ${\mathcal K}_b$ from $F$.
Such a transformation can be very useful in applications, since it allows
to reduce the number of parameters to the minimal. Therefore, in the following,
we aim at the explicit \emph{computation} of $\phi$.

Since the tangent space  has finite codimension, we get that for every $g\in$ $\mathcal U$ there exist $\Gamma$-invariant germs $\mathcal{Q}_i(x,y)$ and real numbers $\mathcal{R}_i$ such that

\begin{equation}\label{inf eq}
g(x,y)=\sum_i \mathcal Q_i(x,y) T_i(x,y) +\mathcal R_1x^2+\mathcal R_2 y^2+\mathcal R_3 x^2y^2
\end{equation}
where ${T_i}$ is a system of generators of $T_f(Gf)$. Equation \eqref{inf eq} is the so called \emph{infinitesimal stability equation} \cite{BrL},
where $\mathcal Q_i$ and $\mathcal R_j$ are, in general, unknown quantities. In the particular case $f=\e_1 x^4+\mu x^2y^2+\e_2y^4$,  a system of generators is given by \cite{Mar}

$$T_1(x,y)=x\frac{\partial f(x,y)}{\partial x}=x(4\e_1x^3+2\mu xy^2),$$

$$T_2(x,y)=y\frac{\partial f(x,y)}{\partial y}=y(2\mu x^2y+4\e_2y^3).$$
Now, suppose that we  are able to solve equation \eqref{inf eq}: then we can  construct the transformation $\phi$ using an iterative algorithm. For simplicity of notation we define $c=(\E,c_1,\dots,c_q)$ where $c_k (k=1,...,q),$ are the set of physical parameters $(a,b)$ in ${\mathcal K}_b$,  $u=(u_1,u_2,u_3)$, $z=(x,y)$ and look for a transformation

\begin{eqnarray}
\phi&:& \mathbb{R}^2\times\mathbb R^{q+1}\rightarrow \mathbb{R}^2\times \mathbb{R}^3 \nonumber \\
      & & (z,c)\rightarrow(\theta(z,c),\rho(c))
\end{eqnarray}
where $\theta: \mathbb{R}^2\times \mathbb{R}^{q+1}\rightarrow \mathbb{R}^2$ is a diffeomorphism which acts as a (parameter depending) coordinate transformation and $\rho: \mathbb{R}^{q+1}\rightarrow\mathbb R^3$ acts as a reparametrisation.

Suppose that we have an algorithm that solves the infinitesimal stability equation modulo terms of order $O(z^d)$, $d\geq2$.
The basic idea is to expand $\theta$ and $\rho$ as formal power series in the parameters $c$ \cite{Kas}:

$$\theta(z,c)=\sum_{j\geq0}\theta_j(z,c),\;\;\;\;\rho(c)=\sum_{j\geq0}\rho_j(c)$$
where $\theta_j$ and $\rho_j$ are homogenous of degree $j$ in $c$. Let us denote

$$\theta^{l}(z,c)\doteq\sum_{i=0}^l\theta_i(z,c)\;\;\;\;\rho^l(c)\doteq\sum_{i=0}^l\rho_i(c)$$
and set $\theta^0(z)= z$, $\rho^0(c)=0$.
Suppose that we are able to compute $\theta$ up to order $l$ in $c$, that is we are able to find $\theta^l$ and $\rho^l$ which solve

\begin{equation}
G(z,c)=F(\theta^{l}(z,c),\rho^l(c))+O(c^{l+1})+O(z^d),
\end{equation}
where $G(z,c)$ is a versal deformation of $f(z)$. Then

\begin{eqnarray}
F(\theta^{l+1},\rho^{l+1})&=&F(\theta^l+\theta_{l+1},\rho^l+\rho_{l+1})= \nonumber\\
&=&F(\theta^l,\rho^l)+D_z F(\theta^l,\rho^l)\theta_{l+1}+D_u F(\theta^l,\rho^l)\rho_{l+1}+\nonumber\\
&+&O(|\theta_{l+1}|^2)+O (|\rho_{l+1}|^2)=\nonumber\\
&=&F(\theta^l,\rho^l)+D_z G(\theta^l,\rho^l)\theta_{l+1}+D_cF(\theta^l,\rho^l)|_{c=0}\cdot\rho_{l+1}
+ O(c^{l+2})\nonumber
\end{eqnarray}
where we obtain the last equality using the estimates $\theta^l(z,c)=z+O(c)$, $\theta_{l+1}(z,c)=O(c^{l+1})$ and $F(z,c)=f(z)+O(c)$.
Thus, we have

\begin{eqnarray} \label{eq inf tr}
G(x,y,c)-F(\theta^l(x,y,c),\rho^{l}(c))&=&\theta_{l+1,1}(x,y,c)\frac{\partial f}{\partial x}+\theta_{l+1,2}(x,y,c)\frac{\partial f}{\partial y}+\rho_{l+1,1}x^2 \nonumber\\
&+&\rho_{l+1,2}y^2+\rho_{l+1,3}x^2y^2+ O(c^{l+2})+O(|x,y|^d).\nn
\end{eqnarray}
This equation has a structure similar to the following one

\begin{eqnarray}
g(x,y,c)&=& x\mathcal Q_{l+1,1}(x,y,c)\frac{\partial f(x,y)}{\partial x}+y\mathcal Q_{l+1,2}(x,y,c)\frac{\partial f(x,y)}{\partial y} +\nonumber\\
&+&\mathcal R_{l+1,1}(c)x^2+\mathcal R_{l+1,2}(c) y^2+\mathcal R_{l+1,3}(c) x^2y^2.\label{eq inf c}
\end{eqnarray}
We can solve \eqref{eq inf c} for the unknowns $\mathcal Q_{l+1,i}$ and $\mathcal R_{l+1,j}$  by equating the coefficients of the monomials $c^{\alpha}=c_1^{\alpha_1}\cdots c_s^{\alpha_s}$ on the left and right hand sides with the condition $\alpha_1+\dots+\alpha_s=l+1$. In such a way we have to solve several equations of the form \eqref{inf eq}.   Thus, if we are able to solve the infinitesimal stability equation, we can find $\mathcal Q_1(x,y,c)$ and $\mathcal Q_2 (x,y,c)$ by solving  \eqref{eq inf c}. If  we  take $\theta_{l+1,1}=xQ_{l+1,1}$, $\theta_{l+1,2}=yQ_{l+1,1}$ and $\rho_{l+1,i}=\mathcal R_{l+1,i}$, we find $\theta$ ad $\rho$ up to order $l+1$ in $c$. In particular we have an explicit expression for the parameters $u_i$ in terms of the $c$, that is

$$u_i=\sum_{j=1}^{l+1}\mathcal R_{j,i}+ O(c^{l+2}),\;\;\;i=1,2,3. $$
An algorithm to solve the infinitesimal stability equation, the so called \emph{division algorithm} \cite{BrL} is presented in subsection \ref{sec:d_alg} below. Using the division algorithm to solve equation \eqref{eq inf tr}
gives the transformation inducing $G$ from $F$. Namely, the following proposition holds

\begin{prop} \label{prop:unf11}
Let ${\mathcal K}_b$ be as in \eqref{Kb}  with central singularity at $\E=b_1=b_2=... =0$
given by $f(x,y)=\e_1x^4+\mu x^2y^2\e_2+\e_2y^4$, $\e_i=\pm1$ for $i=1,2$  and $\mu^2\neq4\e_1\e_2$. There exists
a diffeomorphism $\theta$ and a reparametrisation $\rho$ such that

\be
{\mathcal K}_b(x,y)=F(\theta(x,y,\E,b_i),\rho(\E,b_i))
\ee
with  $\theta(x,y,0)=(x,y)$, $\rho(0,\dots,0)=(0,0,0)$ and

$$F(x,y,u)=f(x,y)+u_1 x^2+u_2 y^2+u_3 x^2y^2.$$
Modulo $\mathcal O(|\E,b_i|^3)+$ $\mathcal O(|x,y|^3)$, the coordinate transformation
$\theta$ reads

\begin{eqnarray}
x&\rightarrow& x \left(1+\e_1\frac{b_5}{4}+\e_1\frac{a_5 \E}{4 }-\frac{3b_5 ^2}{32}-\frac{3 a_5 b_5 \E}{16}-\frac{3a_5^2 \E^2}{32}  \right) \label{eq_1tr11} \\
y&\rightarrow& y \left(1+\e_2\frac{b_7}{4 }+\e_2\frac{a_7 \E}{4 }-\frac{3 b_7^2}{32 }-\frac{3 a_7 b_7 \E}{16 }-\frac{3 a_7^2 \E^2}{32 } \right) \label{eq_2tr11}
\end{eqnarray}
and, modulo $\mathcal O(|\E,b_i|^3)$ the reparametrisation $\rho$ is given by

\begin{eqnarray}
u_1&=&b_2+\left(a_1 +b_1-\e_1\frac{a_5b_2 }{2 }-\e_2\frac{a_1b_5 }{2}\right)\E -\e_1\frac{b_2b_5}{2 }+\left(a_2 -\e_1\frac{a_1 a_5 }{2 }\right)\E^2
\label{u1_gen11}\\
u_2&=& b_4+\left(a_3+b_3 -\e_2\frac{a_7 b_4}{2 }-\e_2\frac{a_3 b_7 }{2}\right)\E-\e_2\frac{b_4 b_7}{2 }+\left(a_4 -\e_2\frac{a_3 a_7 }{2}\right)\E^2
\label{u2_gen11}
\ea

\ba
u_3&=&b_6-\e_1\frac{b_5b_6}{2 }-\e_2\frac{b_6b_7}{2 }+\frac12\left(\frac{3b_5^2 }{4}
-\e_1b_5 +\frac{3 b_7^2 }{4 }-\e_2b_7 +\e_1\e_2\frac{b_5b_7  }{2 }\right)\mu\nn
&+&\left[a_6-\e_1\frac{a_6 b_5}{2 }-\e_1\frac{a_5b_6}{2 }-\e_2\frac{a_7 b_6}{2}-\e_2\frac{a_6b_7}{2 }\right.\nn
&+&\left.\left(\frac{3 a_5b_5 }{8 }-\e_1\frac{a_5 }{2 }+\frac{3 a_7 b_7 }{8 }-\e_2\frac{a_7}{2}+\e_1\e_2\frac{a_7 b_5}{4}+\e_1\e_2\frac{a_5b_7 }{4  }\right)\mu\right]\E\nn
&-&\left[\e_1\frac{a_5a_6}{2 }+\e_2\frac{a_6a_7}{2}-\left(\frac{3 a_5^2 }{8}-\frac{3 a_7^2  }{8 }-\e_1\e_2\frac{a_5 a_7 }{4}\right)\mu\right]\E^2.\label{u3_gen11}
\end{eqnarray}
\end{prop}
\noindent\emph{Proof}.
For $\mu^2\neq4\e_1\e_2$,   since $F$ is a versal deformation of the germ
$\e_1x^4+\mu x^2y^2+\e_2y^4$ the existence of $\theta$ and $\rho$ follows trivially.
By applying the iterative procedure described above to compute  $\theta$ and $\rho$, at each step
we have to solve an equation of type \eqref{eq inf c}. This can be done by exploiting the division algorithm
described in section \eqref{sec:d_alg}. In general, we need to know
 $G$  up to order $L+3$ in order to compute $\theta$
only up to degree $L$ since the first derivatives of the singularity are of degree $3$. Similarly, in order to fix $\rho$, it suffices to know $G$ up to degree four in $(x,y)$ since the maximum degree of the deformation directions (namely $x^2$, $y^2$ and $x^2y^2$) associated to $\rho_1,\dots$, $\rho_3$ is four. Therefore, for ${\mathcal K}_b$ as in \eqref{Kb} the computation can be done  up to and including terms of the first order in $(x,y)$
and the second order in the parameters $(\E,b_i)$ for $\theta$ and  up to and including
$O(|\E,b_i|^2)$ for  $\rho$. With a little computer algebra we obtain the transformations \eqref{eq_1tr11}--\eqref{eq_2tr11} and \eqref{u1_gen11}--\eqref{u3_gen11}. \finedim

\subsection{Solving the infinitesimal stability equation}\label{sec:inf_eq}
We have  seen in the previous section how to construct a transformation inducing $\eqref{Kb}$ from the universal deformation \eqref{F_uni}. Our method is based on the hypothesis that we are able to solve the infinitesimal stability equation \eqref{inf eq} up to a certain  order in the variables $(x,y)$. In this section we present an algorithm to solve this equation. We take the basic ideas from \cite{Br1:2, BrL}.

Let us define $\Sigma^{\Gamma}$ the finite-dimensional vector space of $\Gamma$-invariant  power series on $\mathbb R^2$ truncated at order $d$. We can identify $\Sigma^{\Gamma}$ with the ring $R^{\Gamma}$ of symmetric polynomial in two variables of maximum degree $d$. Let us denote by $z^{ \gamma}$ a monomial in $R^{\Gamma}$ of total degree $\gamma$, that is $z^{ \gamma}=x^{ \gamma_1}y^{ \gamma_2}$, where $1\leq\gamma_1+\gamma_2=\gamma\leq d$.  We can choose an ordering $ \prec $ for monomials in $R$ such that $z^{ \alpha} \prec z^{ \beta}$ if either the total degree of $z^{\alpha}$ is smaller than the total degree of $z^{ \beta}$, or the degree are equal but $z^{ \alpha}$ precedes $z^{ \beta}$ in lexicographic ordering. For example $xy\prec y^2$ since $xy \prec yy$.
\begin{Def}
Let $f$ be a polynomial in $R$.
\begin{description}
  \item [i)] $MM(f)$ is the minimal monomial occurring in $f$ with respect to the monomial ordering described above;
  \item [ii)] $MC(f)$ is the coefficient associated to $MM(f)$;
  \item [iii)]$MT(f)$ is the term associated to $MM(f)$, that is $MT(f)=MC(f) MM(f)$;
  \item [iv)] A monomial $z^{\alpha}$ is said to divide a monomial $z^{\beta}$ if $\beta-\alpha$ is a vector with non-negative entries, then $z^{\beta}/z^{\alpha}=z^{\beta-\alpha}$.
\end{description}
\end{Def}
  If $I={f_1,\dots,f_j}$ is a set of polynomials in $R^{\Gamma}$, we denote by $\langle I \rangle$ the ideal generated by $I$ in $R^{\Gamma}$. The basic idea of the algorithm is to solve the infinitesimal stability equation \eqref{inf eq} through several divisions of  the polynomial $f$ in the ring $R^{\Gamma}$ by the ideal $T$ generated by $\{MM(T_i)\}$, where $\{T_i\}$ is a set of generators of the tangent space to the germ orbit we have described in the previous section. However, in general, the remainder of such a division is not unique. We  need a set of generators for the ideal $T$ which makes the output of such a division unique. This can be done if we choose as a system of generators for $T$ a \emph{Gr\"obner basis} for $T$ with respect to the monomial ordering we have described above. In fact, we recall that a Gr\"obner basis is, by definition, a set of generators for a given $\langle I \rangle$ such that multivariate division of any polynomial in the polynomial ring $R^{\Gamma}$ gives a unique remainder.

  Now, we are ready to present the algorithm.

  \subsection{Division algorithm}\label{sec:d_alg}
  
  Input: integer $d$, power series $f$ truncated at degree $d$, $\{g_1,\dots,g_k\}$ Gr\"obner basis for the ideal $T$.\\
  
  \noindent
  Output: power series $r,q_1,\dots,q_k$ truncated at degree $d$ such that
  $$f=\sum_{i=1}^k q_i g_i +r\;\hbox{ modulo terms of degree } d \hbox{ and highter}.$$

  \noindent
  Algorithm:\\
  \newline
  $h\leftarrow g$ \\
  Reduce $h$ modulo terms of degree $d$ and higher\\
  $r\leftarrow0$\\
  $q_i\leftarrow0$\\
  \textbf{While} $h\neq0$ do

  \textbf{If} $MM(g_i)|MM(h)$ for some i, then\\
  $q_i\leftarrow q_i+MT(h)/MT(g_i)$\\
  $\leftarrow h-(MT(h)/MT(g_i))g_i$\\
  Reduce $h$ modulo terms of degree $d$ or higher

  \textbf{Else}\\
  $r\leftarrow r+MT(h)$\\
  $h\leftarrow h-MT(h)$

  \textbf{End if}\\
  \textbf{End while}.
\newline

Now we have to keep in mind  that we are working in the ring of {\it symmetric} polynomials, thus  we have to make sure  that the output of the division algorithm  respects the $\Gamma$ invariance. In the case we are studying  this is easy to check. In fact, if $\Gamma=\Z_2\times\Z_2$   a polynomial in $R^{\Gamma}$ must be of even degree both in $x$ and $y$. On the other hand, if we consider the germ function $g=\e_1x^4+\mu x^2y^2+\e_2 y^4$, we know that the corresponding invariant tangent space $T$ is generated by $\{2x(2\e_1x^3+2\mu xy^2), 2y(2\mu x^2y+2\e_2y^3)\}$ and a Gr\"obner basis for the ideal $T$ is $GB=\left\{2\e_1 x^4+\mu x^2y^2,2\e_2 x^4+2\mu x^2 y^2,y^6\right\}$ (see e.g. \cite{Br1:2}). Thus, at every step the division algorithm is nothing else but a division between monomials of even degree in both variables. This implies that the outputs of the algorithm are necessarily polynomials of even degree both in $x$ and $y$ an so they respect the $\Gamma$ invariance. In other cases it could be not so easy and the  algorithm must be modified.

\section{Bifurcation curves}\label{sec:bifu}
From section \ref{sec:sing_tr} we know that, if condition \eqref{condition:determinacy} is satisfied,

\be\label{ugerm}
F(x,y;u_1,u_2,u_3)=\e_1 x^4 +(\mu+u_3) x^2 y^2+\e_2y^4+u_1x^2+u_2y^2
\ee
is a universal deformation of $f(x,y)$.
Therefore, there exists a coordinate transformation which induces ${\mathcal K}_b$ from $F$. Such
a transformation can be found by exploiting the algorithm described in the previous section and is given in proposition \ref{prop:unf11}. The phase flows of the corresponding Hamiltonian vector fields being equivalent allows us to deduce the bifurcation sequence and the corresponding energy critical values of the original system from the bifurcation analysis of the simple function \eqref{ugerm}.

Let us now examine the possible inequivalent cases by considering the combinations of the signs of $\e_1$ and $\e_2$.

\subsection{$\bbm[\e_1\e_2=1]$}
The fixed points  of the function \eqref{ugerm}
are given by

\ba
(0,0),\;\left(\pm\sqrt{\frac{-\e_1u_1}{2}},0\right),\; \left(0,\pm\sqrt{\frac{-\e_2u_2}{2}}\right),\label{FP111}\\
 \left(\pm\frac{\sqrt{-u_2-\frac{ \e_2 (-\alpha u_1+2 \e_1u_2)}{\alpha^2-4 \e_1\e_2}}}{\sqrt{\alpha}},\pm \frac{\sqrt{-\alpha u_1+2 \e_1 u_2}}{\sqrt{\alpha^2-4 \e_1\e_2}}\right)\label{FP112}
\ea
where $\alpha=\mu+u_3$. In the case $\e_1\e_2=1$, the corresponding bifurcation curves in the parameter space
are given by $u_1=0$, $u_2=0$, $2u_2+\alpha u_1=0$ and $\alpha u_2+2u_1=0$. Using the  parameters
$u_i$ found in \eqref{u1_gen11}, \eqref{u2_gen11} and \eqref{u3_gen11}, we are able to express these bifurcation curves
 in terms of the the detuning parameter $\d$ and the distinguished parameter $\E$. Namely, we have the following proposition
\begin{prop}\label{prop:bif11}
In the planar system ${\mathcal K}_b$ of proposition \ref{prop:Kb}, bifurcations occur along the following curves in the $(\d,\E)$
plane:

\ba
\E=\E_{1I}&\doteq&\frac{\delta }{2 (2 A-B-6 C)} \label{en_1u}\nn\\
&+&\frac{\left(136 A^2+34 B^2+280 B C+456 C^2-8 A (17 B+70 C)+15 v_{2,4}-45 v_{0,6}\right) \delta ^2}{48 (2 A-B-6 C)^3}\nn
\E=\E_{1L}&\doteq&\frac{\delta }{2 (2 A-B-2 C)} \label{en_1l}\nn\\
&+&\frac{\left(136 A^2+34 B^2+104 B C+72 C^2-8 A (17 B+26 C)+3 v_{2,4}-45 v_{0,6}\right) \delta ^2}{48 (2 A-B-2 C)^3}\nn
\E=\E_{2I}&\doteq&-\frac{\delta }{2 (2 A+B-6 C)}\label{en_2u}\nn\\
&-&\frac{\left(56 A^2+14 B^2+8 A (7 B-74 C)-296 B C+1272 C^2+45 v_{6,0}-15 v_{4,2}\right) \delta ^2}{48 (2 A+B-6 C)^3}\nn
\E=\E_{2L}&\doteq&-\frac{\delta }{2 (2 A+B-2 C)} \label{en_2l}\nn\\
&-&\frac{\left(56 A^2+14 B^2+8 A (7 B-22 C)-88 B C+120 C^2+45 v_{6,0}-3 v_{4,2}\right) \delta ^2}{48 (2 A+B-2 C)^3}\nonumber
\ea
where terms $O(\d^3)$ are neglected.
\end{prop}
\begin{oss}
The fixed points of the planar system ${\mathcal K}_b$ correspond to fixed points for the $1${\rm DOF} Hamiltonian
$\mathcal K$ only if they occur inside the singular circle, cfr remark \ref{oss:singular_circle}. Moreover, the distinguished parameter $\E$ is not negative, therefore the previous curves determine bifurcations
 for the $1${\rm DOF}  system defined by $\mathcal K$ only for those values of the coefficients
and of the detuning parameter which makes (at least) the first order terms non-negative (Arnold ``tongues'').
\end{oss}

\noindent In the following we clarify how the bifurcation curves given in proposition \ref{prop:bif11} have to
be interpreted in terms of the $1$DOF system.

\subsubsection{$\bbm[\e_1=\e_2=-1]$}

To fix the ideas, let us consider the case $C>0$ and $\e_1=\e_2=-1$, which corresponds to $A-3C<A-C<0$, and let us assume that the detuning parameter is not positive.

\begin{oss}\label{oss:DET}
Notice that there is no loss of generality in assuming $\d\leq0$ (i.e. $\omega_1\leq\omega_2$). If in the original phase space we exchange the axes, namely we perform the transformation

\begin{equation} \label{tr_d}
R_1:\; x_1\rightarrow x_2,\;\;\;\;\; x_2\rightarrow x_1,\;\;\;\;\; p_1\rightarrow p_2,\;\;\;\;\;p_2\rightarrow p_1
\end{equation}
the Hamiltonian takes the form

$${\mathcal H'}=\frac12\omega_2(p_1^2+x_1^2)+\frac12\omega_1(p_2^2+x_2^2)+v_{04} x_1^4+...$$
The detuning parameter becomes $\d=\frac{\omega_2}{\omega_1}-1$, which is opposite in sign with respect to the definition \eqref{det11}. Thus, by applying the transformation \eqref{tr_d}, the case $\d>0$ can be treated straightforwardly starting from $\d<0$.
\end{oss}

\begin{figure}[htbp]
\includegraphics[angle=90,angle=90,angle=90,height=9.5cm,keepaspectratio]{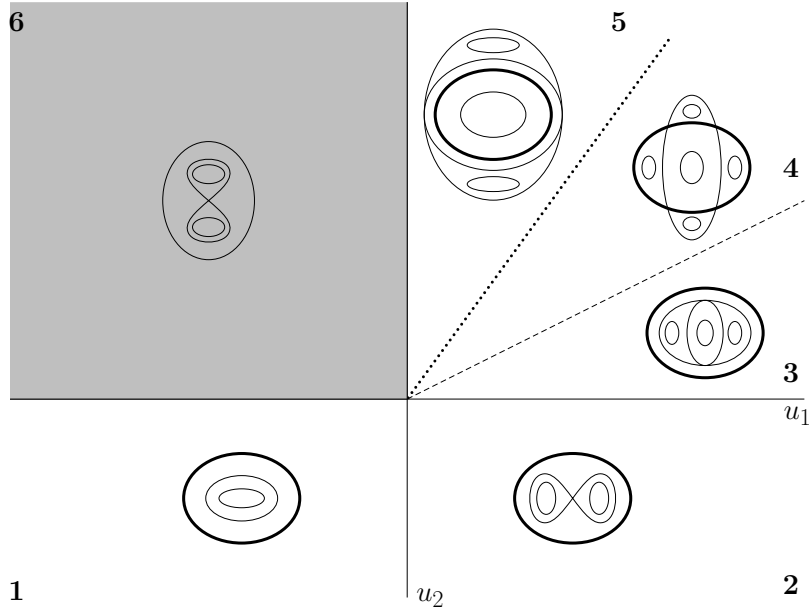}
\caption{\small{Bifurcation diagram in case $A-3C<A-C<0$ and $\d<0$.}}
\label{fig:bifg1}
\end{figure}

\noindent
In this case the deformation $F$ becomes
\be
\mathcal F(x,y)=-x^4+\mu x^2y^2-y^4+u_1x^2+u_2 y^2+u_3x^2y^2.
\ee
The critical points of the planar system are therefore given by \eqref{FP111}--\eqref{FP112} with $\e_1=\e_2=-1$. The fixed points
\be\label{in_loop}
\left(\pm\sqrt{\frac{u_1}{2}},0\right),\quad \left(0,\pm\sqrt{\frac{u_2}{2}}\right)
\ee
bifurcate from the origin when
 $u_1=0$ and $u_2=0$. These critical values of the unfolding parameters respectively determine the bifurcation curves
 \eqref{en_1u} and \eqref{en_1l}.
For $C>0$ and $\d\leq0$, these critical values  correspond to physical acceptable values  if respectively $B>2(A-3C)$ and $B>2(A-C)$. Furthermore, for $\E\approx0$, both $u_1$ and $u_2$ are negative
and $\E_{1I}<\E_{1L}$. Thus, the bifurcations of fixed points \eqref{in_loop} occur according to the diagram given in fig.\ref{fig:bifg1},
from frame $1$ to $3$. The gray zone corresponds to not acceptable values of the parameters. Concerning the critical points

\be
\left( \pm\frac{\sqrt{u_1+\frac{\alpha (-\alpha u_1-2u_2)}{-4+\alpha^2}}}{\sqrt{2}}, \pm \frac{\sqrt{-\alpha u_1-2u_2}}{\sqrt{-4+\alpha^2}}\right), \label{p_xax}
\ee
they determine the bifurcation lines (respectively, the dashed and dotted lines in fig.\ref{fig:bifg1})

\be\label{x_ax_ccurve}
\alpha u_2=-2u_1 , \; 2u_2=-\alpha u_1.
\ee
The expressions of these critical curves in the $(\d,\E)$ plane are given in \eqref{en_2u} and \eqref{en_2l}.
\begin{oss}\label{oss:alpha_ord}
The reduced system comes from a normalization procedure truncated to the fourth order in $\e$, in which
 both $\E$ and $\d$ are assumed to be of second order. Therefore, in the computation of \eqref{en_2u} and \eqref{en_2l} from \eqref{x_ax_ccurve} we retain
in $\alpha$ only terms $O(|\E,\d|)$, since $\alpha$ has to multiply $x^2y^2$ which is a fourth order term.
\end{oss}

\noindent
The critical curves \eqref{en_2u} and \eqref{en_2l} correspond  to acceptable values for  $B>2(C-A)$ and $B>2(3C-A)$.
However, a little computer algebra shows that the critical points \eqref{p_xax} fall on the singular circle \eqref{sing_circ1} (frame $4$ in fig.\ref{fig:bifg1}; the marked circle represents the singular circle), therefore in correspondence of these points the coordinate transformation \eqref{ccoord} is not invertible.
On the other hand,  the fixed points \eqref{in_loop}
could fall on the limit circle, too. At  first order in the deformation parameters, this happens for

\be\label{in_loop_c}
\frac{u_1}{\sqrt{3C-A}}=4\E , \; \frac{u_2}{\sqrt{C-A}}=4\E.
\ee
Solving equations \eqref{in_loop_c} gives the first order term in the detuning parameter of expressions \eqref{en_2u} and \eqref{en_2l}.
This suggests that the critical curves \eqref{en_2u} and \eqref{en_2l} do not determine the bifurcation of new fixed points
for the reduced system defined by \eqref{KER11gen}, but rather the disappearance of  fixed points \eqref{in_loop}.
To verify this statement, we  operate a different planar reduction, according to

\be \label{ccoord2}
\left\{
  \begin{array}{ll}
    x'= & \sqrt{2(\E-J)}\cos\psi,\\
    y'= & \sqrt{2(\E-J)}\sin\psi.
  \end{array}
\right.
\ee
In these coordinates the singularity at $J_2=0$ is removed and we have a singular
circle for $J_1=0$. Proceeding as in the previous section we get the  universal deformation

\be\label{ugerm2}
\mathcal F'(x',y')=-x'^4+(\mu+u'_3) x'^2 y^2-y'^4+u_1'x'^2+u_2'y'^2,
\ee
where the expressions of the deformation parameters are still determined by proposition \ref{prop:unf11}, but the values
of coefficients $a_i$ and parameters $b_i$ change in view of \eqref{ccoord2}. They are listed in appendix
\ref{app:11cat}.

The bifurcation diagram of \eqref{ugerm2} in the $(u_1',u_2')$ plane is still given by fig.\ref{fig:bifg1}.
However, since both $u_1'$ and $u_2'$ turn out to be positive for $\d\leq0$ and $\E\approx0$, in the $(u_1',u_2')$ plane the bifurcation diagram should be read clockwise from $3$ to $1$.
Solving $u_1'=0$ and $u_2'=0$ we find  the critical curves \eqref{en_2u} and \eqref{en_2l}, which therefore must determine the disappearance of
 fixed points \eqref{in_loop} for the reduced Hamiltonian \eqref{KER11gen}, as we claimed.

\begin{oss}
The bifurcation analysis of the reduced system has been performed by assuming $C>0$. For $C<0$ the bifurcation diagram of the germ \eqref{ugerm} remains the same given in fig.\ref{fig:bifg1}. However, since the distinguished parameter must be non-negative and now we have $\E_{1L}<\E_{1I}$, the physical unacceptable zone would be given by panel $2$ and the diagram should be read clockwise starting from frame $1$.
\end{oss}

\noindent Finally, we obtain the following proposition (here and in the following we denote with $(kK), k=1,2, K=I,i,L,\ell$, a given bifurcation: the digit 1 or 2 denotes the normal mode from (to) which the fixed point originates (or annihilates); the letter denotes the bifurcating family ($I$, inclined, stable; $i$, unstable; $L$, loop, stable; $\ell$, unstable).

\begin{prop}\label{prop:mm}
Let us consider the $1${\rm DOF} system $\mathcal K$ defined by \eqref{KER11gen}, with $C\neq0$, $A-3C<0$, $A-C<0$
and non-positive  detuning parameter $\d$. For sufficiently small values of $|\d|$ the following statements hold:\\
\newline
For $C>0$,
\begin{description}
  \item[]i) if $B>2(A-3C)$: a pitchfork bifurcation (a pair of stable fixed points) appears at
  \be \E=\E_{1I} \;\; (1I); \label{1i}\ee
  \item[]ii) if $B>2(A-C)$: a second pitchfork bifurcation (a pair of unstable fixed points) appears at
  \be \E=\E_{1L} \;\; (1\ell); \label{1L}\ee
  \item[]iii) if $B>2(C-A)$: anti-pitchfork bifurcation (the pair of unstable fixed points disappears) at
  \be \E=\E_{2L}\;\; (2 \ell); \label{2L}\ee
    \item[]iv) if $B>2(3C-A)$: a second anti-pitchfork bifurcation (the pair of stable fixed points disappears) at
  \be \E=\E_{2I}\;\; (2I); \label{2i}\ee
    \end{description}
For $C<0$ the bifurcations listed above occur, if the corresponding conditions on $B$ are satisfied,
 but in the different sequence given by $(1L)-(1i)-(2i)-(2L)$.
\end{prop}

\subsubsection{$\bbm[\e_1=\e_2=1]$}
The case $\e_1=\e_2=1$ follows similarly through the bifurcation analysis of
$$-\mathcal  F(x,y)=x^4+(\tilde \mu +\tilde u_3)x^2y^2+y^4+\tilde u_1 x^2+\tilde u_2 y^2$$
where $\tilde\mu=-\mu$, $\tilde u_i=-u_i$, for $i=1,2,3$.
We attain the following proposition:
\begin{prop}\label{prop:pp}
Let us consider the $1${\rm DOF} system defined by $\mathcal K$, with non-positive and sufficiently small detuning parameter, $C\neq0$, $A-3C>0$ and
 $A-C>0$.
\begin{description}
      \item[] If $C>0$ and conditions on $B$ are satisfied in order to give positive values for the energy thresholds, the full bifurcation sequence is given by $(2L)-(2i)-(1i)-(1L)$;
   \item[]if $C<0$ and conditions on $B$ are satisfied in order to give positive values for the energy thresholds, the full bifurcation sequence is given by $(2I)-(2\ell)-(1\ell)-(1I)$.
 \end{description}
\end{prop}

\subsection{$\bbm[\e_1\e_2=-1]$}

\subsubsection{$\bbm[\e_1=-\e_2=-1]$}
\begin{figure}[htbp]
\includegraphics[angle=90,angle=90,angle=90,height=9.5cm,keepaspectratio]{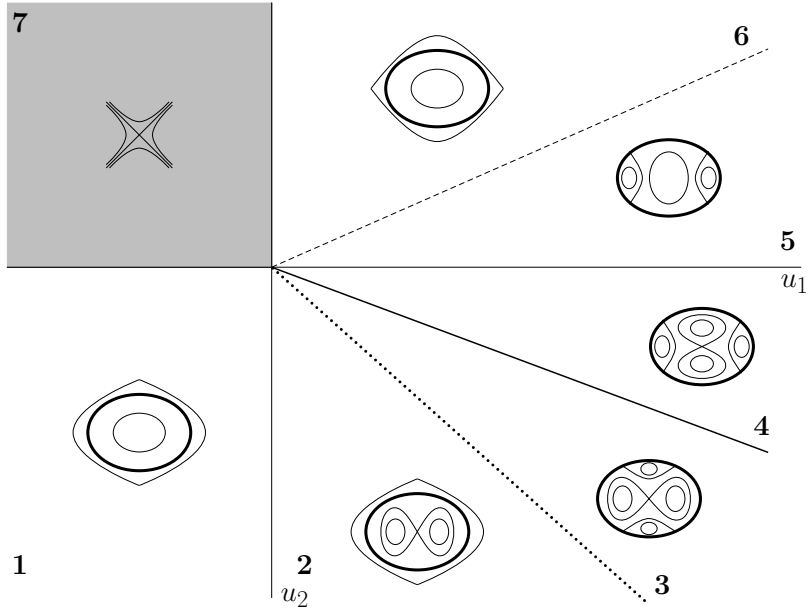}
\caption{\small{Bifurcation diagram for $A-2C<0<A-C$ and $\d<0$.}}
\label{fig:bif2}
\end{figure}

This case corresponds to $A-3C<0$ and $A-C>0$ and
the versal unfolding $F$ turns into

\be\label{unfmp}
\mathcal G(x,y) =-x^4+(\mu +u_3)x^2y^2+y^4+u_1x^2+u_2y^2
\ee
To fix the ideas, let us assume that $A-2C<0$ so that $\mu<0$.
With $\alpha=\mu+u_3$, the critical points of \eqref{unfmp} are then given by $(0,0)$ and

\be\label{cpmp1}
\left(\pm\sqrt{\frac{u_1}{2}},0\right),\;\; \left(0,\pm\sqrt{\frac{-u_2}{2}}\right),
\ee
\be\label{cpmp2}
\left( \pm\frac{\sqrt{u_1+\frac{\alpha (-\alpha u_1-2u_2)}{4+\alpha^2}}}{\sqrt{2}}, \pm \frac{\sqrt{-\alpha u_1-2u_2}}{\sqrt{4+\alpha^2}}\right).
\ee
As we can see in fig.\ref{fig:bif2}, the bifurcation diagram of the system is now quite different from the previous one. Again, we are interested in finding bifurcation curves in the $(\d,\E)$ plane for the one degree of freedom system defined by \eqref{KER11gen}.
Thus, we limit ourselves to consider what happens inside the singular circle \eqref{sing_circ1}, which is marked with a darker line in fig.\ref{fig:bif2}. For $\d\leq0$ and small values of the distinguished parameter, we have $u_1$ and $u_2$ both negative.  The physical unacceptable zone is now given by frame $7$. Thus, the bifurcation sequence has to be read counter-clockwise starting from frame $1$. Therefore the planar system exhibits the first bifurcation at $u_1=0$. The corresponding  bifurcation for the $1${\rm DOF} system defined by \eqref{KER11gen}  occurs for $\E=\E_{1I}$, which is acceptable   only if $B>2(A-3C)$. In frame $3$ we see the appearance of two stable fixed points inside  and four unstable points on the singular circle. By using coordinate transformation \eqref{ccoord2}, we can easily check that,  if $B>2(C-A)$, the corresponding threshold value for the distinguished parameter  is given by \eqref{en_2l} and determines the bifurcation of two stable fixed point for $\mathcal K$. For

\be\label{c_GB}
u_2=-\frac12\left(\alpha +\sqrt{\alpha^2+4}\right)u_1
\ee
(marked line in fig.\ref{fig:bif2} separating panels $3$ and $4$) a \emph{global bifurcation} occurs.
The corresponding threshold value for the distinguished parameter is given by

\be\label{en_gb}
\E=\E_{GB}\doteq-\frac{\d}{2B}+ O(\d^2)
\ee
which is acceptable  if $B>0$. Notice that since $\alpha$ multiplies a fourth order term, we can consider
\eqref{c_GB} only up to the first order in $|\d,\E|$, cfr remark \ref{oss:alpha_ord}. Therefore, we are able to compute the critical curve
\eqref{en_gb} only to the first order in the detuning parameter.
Then, if $B>2(A-C)$, we can pass through $u_2=0$ for $\E=\E_{1L}$  and if $B>2(3C-A)$ a further bifurcation occurs when passing through $u_1=0$; the corresponding threshold value for the distinguished parameter is given by  \eqref{en_2u}.

The case $\mu>0$ follows similarly through the bifurcation analysis
of
$$-\mathcal G(y,x)=-x^4+(\tilde \mu+\tilde u_3) x^2y^2+y^4+\tilde u_1x^2+\tilde u_2y^2$$
where  $\tilde \mu=-\mu<0$, $\tilde u_j=-u_j$ for $j=1,2,3$.
Finally we have the following proposition:

\begin{prop}\label{prop:mp}
Let us consider the $1${\rm DOF} system $\mathcal K$ defined by \eqref{KER11gen} with non-positive and sufficiently small detuning parameter, $C\neq0$, $A-3C<0$ and $A-C>0$:

For $A-2C<0<A-C$, if conditions on $B$ are satisfied in order to get positive values of the energy thresholds, bifurcations occur along the curves \eqref{en_1u}-\eqref{en_2l} in the sequence $(1I)-(2L)-(1L)-(2I)$. Furthermore,
a global bifurcation might occur between $(2L)$ and $(1L)$ if $B>0$ at
\be \E=\E_{GB} \;\; (GB) \ee
with $\E_{2L}<\E_{GB}<\E_{1L}$.

For $0<A-2C<A-C$, if conditions on $B$ are satisfied in order to give positive values for the energy thresholds, the full bifurcation sequence is given by $(2L)-(1I)-(GB)-(2I)-(1L)$.
\end{prop}

\subsubsection{The degenerate case $A=3C$}

It remains to analyze the case $\mu=0$, corresponding  to the central singularity
$y^4-x^4$.
In this case  \eqref{ugerm} turns into
\be
\mathcal F(x,y)=-x^4+y^4+ u_1x^2+u_2 y^2+u_3 x^2y^2.
\ee
The critical points remain the same given in \eqref{cpmp1} and \eqref{cpmp2}, but we now have $\alpha=u_3$. The bifurcation curves
are therefore given by

\be\label{bcurve_muzero}
u_1=0,\;\;\;\;\; u_2=0
\ee
and

\ba\label{ccbcurve_muzero}
2u_1&=&u_3u_2,\label{ccbcurve_muzero1}  \\
2u_2&=&-u_1 u_3.\label{ccbcurve_muzero2}
\ea
Solving \eqref{bcurve_muzero}, we find the critical values $\E=\E_{1I}$ and $\E=\E_{1L}$,
which respectively turns out to satisfy also \eqref{ccbcurve_muzero1} and \eqref{ccbcurve_muzero2}. Thus, we get

\be
\E_{1I}=\E_{2L}\;\;\;\; \hbox{and} \;\;\;\; \E_{2I}=\E_{1L}.
\ee
For the $1$DOF system defined by \eqref{KER11gen}, this implies that the  critical points in \eqref{cpmp1}
appear and disappear simultaneously. Furthermore, a global bifurcation occurs for

\be
u_2=-\frac12\left(u_3 +\sqrt{u_3^2+4}\right)u_1
\ee
giving the critical curve \eqref{en_gb}. Summarizing, the following proposition holds:
\begin{prop}\label{prop:muzero}
In the $1${\rm DOF} system $\mathcal K$ defined by \eqref{KER11gen}, for $A=2C>0$ and non-positive sufficiently small values of the detuning parameter, we have
\begin{description}
   \item[]i) if $B>-2C$ two pitchfork bifurcations occur concurrently (two pairs of stable fixed points
   appear) at
   $$\E=\E_{1I}=\E_{2L};$$
   \item[]ii) if $B>0$ a global bifurcation occurs at
   $$\E=\E_{GB};$$
   \item[]iii) if $B>2C$ two anti-pitchfork bifurcations occur concurrently (the two pairs of stable fixed points disappear) at
   $$\E=\E_{1L}=\E_{2I}.$$
 \end{description}
\end{prop}

\subsubsection{$\bbm[\e_1=-\e_2=1]$}

This last sub-case (treated in \cite{Br1:1} as a realization of the $\mathbb{Z}_2$-symmetric 1:1 resonance in the ``spring-pendulum'') follows similarly through the bifurcation analysis of
$$-{\mathcal G}(x,y)= x^4+(\tilde \mu+ \tilde u_3) x^2y^2-y^4+\tilde u_1x^2+\tilde u_2y^2$$
with $\tilde \mu=-\mu$, $\tilde u_i=-u_i$, for $i=1,2,3$. Therefore,  the following proposition holds:

\begin{prop}\label{prop:pm}
Let us consider the $1${\rm DOF} system $\mathcal K$ defined by \eqref{KER11gen}, with $C\neq0$, $A-3C>0$ and $A-C<0$. For non-positive and sufficiently small detuning parameter, bifurcations might occur along the curves \eqref{en_1u}--\eqref{en_2l} and \eqref{en_gb} in agreement with the statements
of propositions \ref{prop:mm}--\ref{prop:muzero}. If conditions on $B$ are satisfied in order to give positive values for the energy thresholds,

for $A-2C<0<A-3C$, the full bifurcation sequence is given by
 $(1L)-(2I)-(GB)-(1I)-(2L)$;

for $0<A-2C<A-3C$
it is given by $(2I)-(1L)-(GB)-(2L)-(1I)$;

for $A=2C$ bifurcations occur according to the statements of proposition \ref{prop:muzero},
but they are reached in the sequence $(1L)-(GB)-(1I)$.
\end{prop}

\section{Implications for the original system}\label{sec:original11}
According to these results, with the versal deformation of this resonance, we know number and nature of the critical points. Including higher orders may shift the positions of the equilibria --
and may be essential for quantitative uses -- but will not alter their number
or stability. The isolated equilibria of the $1$DOF system defined by \eqref{KER11gen} correspond to relative equilibria for the original system \eqref{Hamiltonian} \cite{DE, M1}. Of course, the results obtained are limited to low energies,
in the neighbourhood of a central equilibrium, but extend to the original system defined by Hamiltonian (1). This statement is based on the fact that the difference between the original Hamiltonian and the normal form (namely, the remainder of the normalization) can be considered a perturbation of the normal form itself \cite{HS, MH, SV}. This remark implies that the results concerning periodic orbits can be extended, by applying the implicit function theorem, to the original system \cite{Henrard,Ku}. On the same ground, iso-energetic KAM theory \cite{Arn63, FHPY} can be used to infer the existence of invariant tori seen as non-resonant tori of the normal form surviving when perturbed by the remainder.

Since we pushed the normalization  up to and
including sixth order terms, the critical curves of proposition \ref{prop:mm} give quantitative predictions on the bifurcation and stability of these periodic orbits in the $(\d,\E)$-plane up to second order in the detuning parameter (since, we recall, it is assumed to be associate to a term of higher-order in the series expansions).

For the coordinate transformation \eqref{ccoord}, the origin in the plane is a fixed point for all values
of the parameters and  represents  the periodic orbit $J_1=0$, namely the normal mode along the $x_2$-axis (the ``short-period'' one, in the reference case $\d<0$). Similarly, if the
planar reduction is performed via \eqref{ccoord2},  we find that for all values of the
parameters, the origin  is a fixed point  again, but it corresponds in this case
to the periodic orbit $J_2=0$, that is the normal mode along the $x_1$-axis (the ``long-period'' one). In the previous section we found  threshold values for the distinguished parameter,  depending on $\d$
and on the coefficients of the system, which determine the bifurcation of these periodic orbits in general position
from the normal modes of the system. However, it would be better to have an expression of the bifurcation curves in the $(\d,E)$-plane, where $E$ is the ``true'' energy of the system. On the long-period axial orbit ($J_2=0$, $J_1=\E$), we have

\ba
\mathcal K&=&\E(1+\d)+(2 A +B) \E^2\nn
&+& \frac{\E^2}{18} \left(-136 A^2 \E-136 A B \E-34 B^2 \E+45 v_{6,0} \E-72 A \delta -36 B \delta \right).\nn
\label{ks11}
\ea
According to the rescaling (\ref{scaling_t}), $E=\omega_2 \mathcal K$, so that equation (\ref{ks11}) can be
used to express the physical energy $E$ in terms of $\E$, namely

\be
E= \omega _2 \left[\E (1+\d) +\left(2 A +B \right)\E^2  \right] .
\ee
Thus up to second order in $\d$, for $\E$ satisfying equations (\ref{en_2u}), (\ref{en_2l}) and $\d$ as defined in \eqref{det11}  we obtain the following threshold values

\ba
E&=&E_{2I}\doteq -\frac{  \omega _2 }{2 (2 A+B-6 C)} \delta\\
&-&\frac{\left(104 A^2+104 A B+26 B^2-1024 A C-512 B C+2136 C^2+45 v_{6,0}-15 v_{4,2}\right)  \omega _2 }{48 (2 A+B-6 C)^3}\delta ^2\nn
E&=&E_{2L}\doteq -\frac{  \omega _2}{2 (2 A+B-2 C)} \delta\\
&+&\frac{\left(104 A^2+104 A B+26 B^2-320 A C-160 B C+216 C^2+45 v_{6,0}-3 v_{4,2}\right) \omega _2}{48 (2 A+B-2 C)^3}  \delta ^2\nonumber
\ea
for the appearance (disappearance) of respectively inclined  and loop orbits from the long-period axial orbit.
They correspond to physically acceptable values, at least for small values of $|\d|$, if

\be\label{phys_conditionsx_2}
2A+B-6C>0,\;\;\;\hbox{ and }\;\;\;2 A+B-2 C>0.
\ee
These conditions are reversed for $\d>0$. A similar argument gives the threshold values for the bifurcations  from the short-period axial orbit.
They are given by

\ba
E&=&E_{1I}\doteq \frac{ \omega _2}{2 (2 A-B-6 C)} \delta \\
&+&\frac{\left(184 A^2-184 A B+46 B^2-704 A C+352 B C+456 C^2+15 v_{2,4}-45 v_{0,6}\right) \omega _2 }{48 (2 A-B-6 C)^3}\delta ^2\nn
E&=&E_{1L}\doteq\frac{ \omega _2}{2 (2 A-B-2 C)} \delta \\
&+&\frac{\left(184 A^2-184 A B+46 B^2-256 A C+128 B C+72 C^2+3 v_{2,4}-45 v_{0,6}\right) \omega _2 }{48 (2 A-B-2 C)^3}\delta ^2\nonumber
\ea
and correspond to physically acceptable values, at least for small values of the detuning parameter, if

\be\label{phys_conditionsx_1}
2A-B-6C<0,\;\;\;\hbox{ and }\;\;\;2A-B-2C<0.
\ee
Finally, the global bifurcation may occur at

\be
E=E_{GB}\doteq-\frac{\omega_2\d}{2B}
\ee
if

\be
B>0.
\ee
The bifurcation sequences of the original system depend on the three coefficients $A,B,C$
according to the statements of propositions \ref{prop:mm}--\ref{prop:pm} as we have obtained them in the previous section.

\appendix
\section{Normal Forms}\label{appendice}

\subsection{Action-angle--like variables}\label{sec:aa}

The terms in the Birkhoff normal form \eqref{aav11} are

\ba
K_0 &=& J_1 + J_2,\\
K_2 &=&\d J_1 + \frac32 v_{4,0} J_1^2+ \frac32 v_{0,4}  J_2^2+ \frac12 v_{2,2} J_1 J_2 \left[2 + \cos(2\phi_1- 2\phi_2) \right],\\
K_4 &=&-\d\left[3 v_{4,0} J_1^2  - v_{2,2} J_1 J_2 \left(2 + \cos(2\phi_1- 2\phi_2)\right)\right],\label{aav11_normal_form}\nn
&+&\frac14 \left(10 v_{6,0}-17 v_{4,0}^2\right)J_1^3 +\frac14 \left(10 v_{0,6}-17 v_{0,4}^2\right)J_2^3\nn
&+&\frac14 J_1^2J_2 \left(6v_{4,2}-\frac94 v_{2,2}^2 - 12 v_{2,2} v_{4,0} +\left(4v_{4,2} - 2 v_{2,2}^2 - 5 v_{2,2} v_{4,0} \right) \cos(2\phi_1- 2\phi_2)\right)\nn
&+&\frac14 J_1J_2^2 \left(6v_{2,4}-\frac94 v_{2,2}^2 - 12 v_{2,2} v_{0,4} +\left(4v_{2,4} - 2 v_{2,2}^2 - 5 v_{2,2} v_{0,4} \right) \cos(2\phi_1- 2\phi_2)\right).\nn
\ea

\subsection{Variables for the first reduction}\label{sec:bb}

With the definitions

\be
A=\frac38 (v_{4,0}+v_{0,4}),\;\;\;
     B=\frac34 (v_{4,0}-v_{0,4}),\;\;\;
C=\frac18 v_{2,2},
\ee
the polynomials in the reduced normal form \eqref{KER11gen} are

\ba
\mathcal A_1(J;\E,\d)&=&(2A-B)\E^2+\left( -4A +2 B +8 C\right)\E J\nn
&+&\d J+4(A-2C)J^2.\\
\mathcal B_1(J;\E,\d)&=& 4C J(\E-J).\\
\mathcal A_2 (J;\E,\d)&=&\frac19 \left(68 A B-68 A^2 -17 B^2 +\frac{45}{2}v_{0,6}\right)\E^3
-8 C \delta \E J \nn
&+&\left(+\frac{68 A B}{3}-\frac{68 A^2}{3}-\frac{17 B^2}{3}+32 A C-48 B C+36 C^2\right.\nn
&+&\left.\frac{3 v_{4,2}}{2}-3v_{2,4} +\frac{15 v_{0,6} }{2}\right)\E J^2+ 2(4C-2 A -  B )\d J^2 \nn
&+&\left(32 B C+\frac{5 v_{6,0}}{2}-\frac{136 A B}{9}-\frac{3v_{4,2}}{2}+\frac{3v_{2,4}}{2}-\frac{5 v_{0,6}}{2}\right)J^3.\\
\mathcal  B_2(J;\E,\tilde \d)&=& -4 C \delta\E J+\left(\frac{40}{3} A C -20 B C +32 C^2 +v_{4,2} -2 v_{2,4}\right)\E J^2\nn
&+&4C\d J^2+\left(\frac{40}{3} B C -v_{4,2}+v_{2,4}\right)J^3.
\ea

\section{List of coefficients and parameters}\label{app:11cat}

In the deformation \eqref{Kb} of proposition \ref{prop:Kb}, if the planar reduction is performed according to \eqref{ccoord}, the coefficients $a_i$ are the following
\ba
a_1&=&\frac{12( B-2A+6 C)}{ \sqrt{|3C-A|}} \nn
a_2&=&\frac{136 A^2-136 A B+34 B^2-272 A C+136 B C-408 C^2+15 v_{2,4}-45 v_{0,6}}{12 \sqrt{|3C-A|}}\nn
a_3&=&\frac{12(B-2 A+2 C)}{ \sqrt{|C-A|}}\nn
a_4&=&\frac{136 A^2-136 A B+34 B^2-112 A C+56 B C-24 C^2+3 v_{2,4}-45 v_{0,6}}{12 \sqrt{|C-A|}}\nonumber
\ea
\ba
a_5&=&\frac{1}{288 (A-3 C)^2}(1632 A^3-544 A^2 (2 B+15 C)+2 A (68 B^2+3264 B C\nn
&+&9 (272 C^2-5 (v_{6,0}+v_{4,2}-3 v_{2,4}+5v_{0,6})))\nn
&+&3 (-136 B^2 C-3 B(1088 C^2-5 (v_{6,0}-v_{4,2}+v_{2,4}-v_{0,6}))\nn
&+&18 C (272 C^2+5 (v_{6,0}+v_{4,2}-3v_{2,4}+5 v_{0,6}))))\nn
a_6&=&\frac{1}{\left(144 |C-A|^{3/2} |3C-A|^{3/2}\right)}\left[-1632 A^4+64 A^3 (17 B+138 C)\right.\nn
&-&2 A^2 \left(68 B^2+3328 B C+9 \left(640 C^2-5 v_{6,0}-3 v_{4,2}+9 v_{2,4}-25 v_{0,6}\right)\right)\nn
&+&A (512 B^2 C-72 C (48 C^2+5 v_{6,0}+2 v_{4,2}-8 v_{2,4}+25 v_{0,6})\nn
&+&B (12608 C^2-45v_{6,0}+27 v_{4,2}-27 v_{2,4}+45 v_{0,6}))\nn
&-&6 C (52 B^2 C+B (1216 C^2-15 v_{6,0}+6 v_{4,2}-6 v_{2,4}+15 v_{0,6})\nn
&-&3 C (432 C^2+5 v_{6,0}+7 v_{4,2}-25v_{2,4}+85v_{0,6}))] \nonumber
\ea
\ba
a_7&=&\frac{1}{288 (A-C)^2}\left(1632 A^3+32 A^2 (34 B-93 C)\right.\nn
&+&2 A \left(68 B^2-1216 B C+528 C^2-225 v_{6,0}+27 v_{4,2}-9 v_{2,4}-45 v_{0,6}\right)\nn
&+&3 \left(-24 B^2 C+B \left(448 C^2+3 (5 v_{6,0}-v_{4,2}+v_{2,4}-5 v_{0,6})\right)\right.\nn
&+&\left.\left.6 C \left(16 C^2+25 v_{6,0}-3v_{4,2}+ v_{2,4}+5 v_{0,6}\right)\right)\right).\nonumber
\ea
The parameters $b_i$ have the following expressions
\ba
b_1&=&-\frac{6 C\d }{\sqrt{|3C-A|}}\nn
b_2&=&\frac{\delta }{2 \sqrt{|3C-A|}}\nn
b_3&=&-\frac{2 C \delta }{\sqrt{|C-A|}}\nn
b_4&=&\frac{\delta }{2 \sqrt{|C-A|}}\nn
b_5&=&\frac{\left(576 A^2+16 A B-3456 A C-48 B C+5184 C^2+45 (v_{6,0}-v_{4,2}+v_{2,4}- v_{0,6})\right) \delta }{576 (A-3 C)^2}\nn
b_6&=&\left(-576 A^3-16 A^2 B+3456 A^2 C+32 A B C-6336 A C^2\right.\nn
&+&48 B C^2+3456 C^3-45 A v_{6,0}+90 C v_{6,0}+27 A v_{4,2}-36 C v_{4,2}-27 A v_{2,4}+36 C v_{2,4}\nn
&+&45 A v_{0,6}-90 Cv_{0,6})\frac{ \delta }{288 |C-A|^{3/2} |3C-A|^{3/2}}\nn
b_7&=&\frac{\left(576 A^2+16 A (B-72 C)+48 B C+9 \left(64 C^2+5 v_{6,0}-v_{4,2}+v_{2,4}-5 v_{0,6}\right)\right) \delta }{576 (A-C)^2}.\nn
\nonumber
\ea
If the planar reduction is performed according to \eqref{ccoord2} the previous coefficients and parameters turn into
\ba
a_1&=&-\frac{12( B+2A-6 C)}{ \sqrt{|3C-A|}} \nn
a_2&=&\frac{136 A^2+136 A B+34 B^2-272 A C-136 B C-408 C^2+15 v_{4,2}-45 v_{6,0}}{12 \sqrt{|3C-A|}}\nonumber
\ea
\ba
a_3&=&-\frac{12(B+2 A-2 C)}{ \sqrt{|C-A|}}\nn
a_4&=&\frac{136 A^2+136 A B+34 B^2-112 A C+56 B C-24 C^2+3 v_{4,2}-45 v_{6,0}}{12 \sqrt{|C-A|}}\nn
a_5&=&\frac{1}{288 (A-3 C)^2}(1632 A^3+544 A^2 (2 B-15 C)+2 A (68 B^2-3264 B C\nn
&+&9 (272 C^2-5 (v_{6,0}-3v_{4,2}+v_{2,4}-v_{0,6})))\nn
&+&3 (-136 B^2 C+3 B(1088 C^2+5 (v_{6,0}-v_{4,2}+v_{2,4}-v_{0,6}))\nn
&+&18 C (272 C^2+5 (v_{6,0}-3v_{4,2}+v_{2,4}+ v_{0,6}))))\nn
a_6&=&\frac{1}{\left(144 |C-A|^{3/2} |3C-A|^{3/2}\right)}\left[-1632 A^4-64 A^3 (17 B-138 C)\right.\nn
&-&2 A^2 \left(68 B^2-3328 B C+9 \left(640 C^2-25 v_{6,0}+9 v_{4,2}-3 v_{2,4}-5 v_{0,6}\right)\right)\nn
&+&A (512 B^2 C-72 C (48 C^2+25 v_{6,0}-8 v_{4,2}+2 v_{2,4}+5 v_{0,6})\nn
&+&B (-12608 C^2-45v_{6,0}+27 v_{4,2}-27 v_{2,4}+45 v_{0,6}))\nn
&+&6 C (-52 B^2 C+B (1216 C^2+15 v_{6,0}-6 v_{4,2}+6 v_{2,4}+5 v_{0,6})\nn
&+&3 C (432 C^2+85 v_{6,0}-25 v_{4,2}+7v_{2,4}+5v_{0,6}))] \nonumber
\ea
\ba
b_1&=&\frac{2 (2 A+B-3 C) \delta }{\sqrt{-A+3 C}}\nn
b_2&=&-\frac{\delta }{2 \sqrt{|3C-A|}}\nn
b_3&=&\frac{2 (2 A+B-C) \delta }{\sqrt{-A+C}}\nn
b_4&=&-\frac{\delta }{2 \sqrt{|C-A|}}\nn
b_5&=&\frac{\left(576 A^2+16 A B-3456 A C-48 B C+5184 C^2+45 (v_{6,0}-v_{4,2}+v_{2,4}- v_{0,6})\right) \delta }{576 (A-3 C)^2}\nn
b_6&=&\left(-576 A^3-16 A^2 B+3456 A^2 C+32 A B C-6336 A C^2\right.\nn
&+&48 B C^2+3456 C^3-45 A v_{6,0}+90 C v_{6,0}+27 A v_{4,2}-36 C v_{4,2}-27 A v_{2,4}+36 C v_{2,4}\nn
&+&45 A v_{0,6}-90 Cv_{0,6})\frac{ \delta }{288 |C-A|^{3/2} |3C-A|^{3/2}}\nn
b_7&=&\frac{\left(576 A^2+16 A (B-72 C)+48 B C+9 \left(64 C^2+5 v_{6,0}-v_{4,2}+v_{2,4}-5 v_{0,6}\right)\right) \delta }{576 (A-C)^2}\nonumber
\ea

\end{document}